\documentclass[aps,prl,twocolumn,showpacs,preprintnumbers]{revtex4}
\usepackage{amsmath,amssymb}
\usepackage{graphicx}
\usepackage{color}
\newcommand\beq{\begin{equation}}
\newcommand\eeq{\end{equation}}
\newcommand\beqa{\begin{eqnarray}}
\newcommand\eeqa{\end{eqnarray}}

\newcommand\F{{\scriptscriptstyle F}}
\newcommand\Hd{{\scriptscriptstyle H}}

\newcommand\T{{\scriptscriptstyle T}}

\newcommand\etc{{\sl etc\/}}

\newcommand\etal{{\sl et al.\/}\ }

\newcommand\ejp{{Eur.\ J.\ Phys.\/}\ }
\newcommand\epjc{{Eur.\ Phys.\ J.\/} C\ }
\newcommand\jhep{{J.\ H.\ E.\ P.\/}\ }
\newcommand\npa{{Nucl.\ Phys.\/} A\ }
\newcommand\npb{{Nucl.\ Phys.\/} B\ }
\newcommand\plb{{Phys.\ Lett.\/} B\ }
\newcommand\zpc{{Z.\ Phys.\/} C\ }

\begin{document}
\title{Fractal structure of near-threshold quarkonium production
 off cold nuclear matter}
\author{Partha Pratim\ \surname{Bhaduri}}
\email{partha.bhaduri@vecc.gov.in}
\affiliation{Variable Energy Cyclotron Center, 1/AF Bidhan Nagar,
 Kolkata 700064, India.}
\author{Sourendu\ \surname{Gupta}}
\email{sgupta@theory.tifr.res.in}
\affiliation{Department of Theoretical Physics, Tata Institute of Fundamental
         Research,\\ Homi Bhabha Road, Mumbai 400005, India.}
\begin{abstract}
We investigate near-threshold production of quarkonium resonances in
cold nuclear matter through a scaling theory with two exponents which
are fixed by existing data on near-threshold $J/\psi$ production in
proton-nucleus collisions.  Interestingly, it seems possible to extend
one of the multifractal dimensions to the production of other mesons
in cold nuclear matter. The scaling theory can be tested and refined
in experiments at the upcoming high-intensity FAIR accelerator complex
in GSI.
\end{abstract}
\pacs{25.40.Ep,14.40.Pq,24.85.+p.\hfill TIFR/TH/13-09}
\maketitle

There are unexplored systematics for the production of quarkonia close to
threshold. For the $J/\psi$, a few experimental studies were carried out
in the days before the understanding of QCD was mature.  Attention shifted
to the high-energy frontier, since perturbative QCD turned out to be the
tool appropriate to that region, and weak-coupling theory was applied
successfully to higher-energy production of quarkonium in pA collisions
\cite{hard,sridhar,maltoni,nelson}.  However, the planned FAIR in GSI
presents a grand opportunity for the study of near-threshold particle
production in cold nuclear matter, and test and refine a scaling theory
which is developed here. This is interesting for several reasons. First,
as we argue here, testing the limits of such a scaling theory shows
where the crossover between hadron and quark descriptions of matter lie.
Second, this measurement can be used to test factorization where it
cannot be proved, and therefore has implications on the understanding of
CP violations. Third, a detailed understanding of cold nuclear effects
in quarkonium production is important to tests of the formation of the
QCD plasma in heavy-ion collisions. Finally, the discovery of scaling
laws, which is our main result, is of fundamental interest since the
exponents can define universality classes across very broad ranges of
physical phenomena.

The basic toolkit which we bring to this study is the modern understanding
that the renormalization group is expressed via scaling. We wish to
ascertain whether there is a dynamical symmetry of cold nuclear matter so
that a small change in the amount of nuclear matter can be compensated for
by a corresponding change in the energy of the probe.  Invariance under
such scaling would manifest itself in the form of exponents which are
eigenvalues of renormalization group transformations \cite{scaling}. These
scaling exponents are also called fractal dimensions or anomalous
dimensions. At high energies they can be computed in perturbation
theory. Near the threshold of particle production, they have to be
discovered in data and understood non-perturbatively.  Discovery of
scaling exponents, which is our main result, conversely implies the
scaling symmetries.

We are interested in the production of a quarkonium state, H, in pA
collisions near the threshold energy $\sqrt S_0$, which is the minimum
energy in the center of mass required to produce H.  We follow the
convention of writing the CM energy, $\sqrt S$, in the equivalent
pp system; for a fixed target configuration this means $S=2M_p(E_b+M_p)$,
where $E_b$ is the beam energy and $M_p$ the proton mass. The threshold
energy, $\sqrt{S_0}=2M_p+M_\Hd$, where $M_\Hd$ is the mass of H.  The total
inclusive cross section, $\sigma$, can be a function of $\sqrt S$,
$M_\Hd$, $M_p$, and the nuclear mass, $M_A$.  Then, a dimensional
argument allows us to write
\beqa
\nonumber
 S_0\sigma &=& f(A,Y,h),\quad{\rm where}\quad
   A=\frac{M_A}{M_p},\\
 &&\;
   Y=\frac12\log\left(\frac S{S_0}\right),\;\;
   h=\frac{M_\Hd}{M_p},
\label{dim}\eeqa
and $f$ is a dimensionless function.  In this definition of $A$, we
neglect the effects of nuclear binding, which are expected to be less
than 1\%, and isospin effects, which could be slightly larger.  We take
masses and branching ratios from \cite{pdg}.

In this paper we report a scaling analysis of $J/\psi$ cross sections
in a dilepton channel in pp and pA collisions from the lowest up to
ISR energies \cite{CERN-PS, IHEP, E331, NA60, NA3, NA38, NA50-400,
NA50-450, E771, E789, E866, HERAB, Aubert, WA39, NA51, ISR}, but not
beyond. Within this data corpus, corrections for kinematic acceptance
limitations of each experiment needed for global analyses are discussed
in \cite{maltoni,ramona}.  We decided to examine $B\sigma$ rather than
$\sigma$, where $B$ is the branching ratio in the dielectron or dimuon
channel. The reason is that over the years the value of $B$ has moved
by more than its error bar. When the inclusive cross section in one of
these dilepton channels is measured, this uncertainty does not affect
the result.  Some experiments correct their data for nuclear effects
according to a formula $A^\alpha$, with $\alpha$ obtained from their
data. We undid this correction, since this is part of our global analysis.

Near threshold the variable $Y$ is small and close to zero. In proton
nucleus scattering, we expect some Fermi motion: even though the CM of
the nucleus is at rest, individual nucleons may be moving. The typical
energy of this movement is of the order of the binding energy per
nucleon \cite{migdal}, and hence comparable to other effects which we
have neglected.  Clearly, Fermi motion can be detected with experiments
close to threshold since the cross section vanishes otherwise.  However,
for $Y>0.1$, the effect can be neglected.

When $\sqrt S$ is large, then one expects to be able to compute
$f(A,Y,h)$ in models inspired by perturbative QCD; this is true whether
$h$ is large and $Y$ small \cite{top} or $h$ is small and $Y$ large
\cite{hard,sridhar,maltoni,nelson}.  An important pre-requisite for
these computations is the factorization of the initial state into parton
distribution functions. If these factorization theorems were valid then
certain kinematic scalings could be expected to hold \cite{helmut} which
are seen to fail for $Y\approx1$ \cite{NA60}.  So, in the near-threshold
region for $J/\psi$ perturbative QCD inspired models for quarkonium
production is not viable.

When $Y$ is small, one expects the cross section to have a Taylor
expansion in $Y$. In pp collisions, where $A=1$, one may then expect the
particularly simple parametrization $f_1(Y,h) = Y^\beta f_1(\widetilde
h)$, where $\beta$ is a scaling dimension and $\widetilde h$ is a scaled
variable. Such a scaling law indeed describes the $A=1$ part of the data
corpus well with
\beq
   Bf_1/S_0 = 2.0\pm0.4\;{\rm nb\/}, \quad
   \beta = 3.20\pm0.26,
\label{ppscale}\eeq
with covariance  of $-0.934$. 

More generally, we can write $f(A,Y,h) = Y^\beta f(\widetilde A,
\widetilde h)$, where the ``renormalized'' variables are taken to have the
form $\widetilde A=A/(y+Y)^\mu$, where $\mu$ is an anomalous dimension
and $y$ is an additive renormalization constant. Then a change from $Y$
to $Y'=\zeta Y$ can be compensated by a change of the target nucleus
from $A$ to $A'=\xi A$ with $\xi^{1/\mu} = (\zeta Y+y)/(Y+y)$. One has
a similar renormalized $\widetilde h(h,Y)$.  A power law $f(\widetilde
A,\widetilde h)=\widetilde A^{\widetilde \alpha} f(\widetilde h)$ then
gives a multifractal exponent $A^{\alpha(Y)}$.  Generally, such a power
law parametrization may be expected when $A\gg1$.  Using the extensive
data taken by NA50 \cite{NA50-400} we find that this is indeed a very
good description of the data for $A>50$. Furthermore it is significantly
less probable that this behaviour extends to all $A$.

\begin{figure}[tp]
\begin{center}
\includegraphics[scale=0.7]{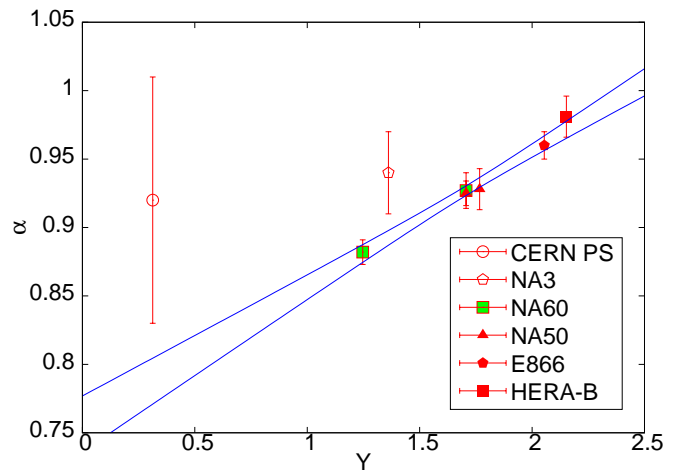}
\end{center}
\caption{The multifractal exponent $\alpha$ for the production of $J/\psi$
 in pA collisions. The band encloses the 68\% confidence limits of the two
 parameter fit of eq.\ (\ref{alpha}).}
\label{fg.alpha}\end{figure}

Moreover, from the data corpus it is clear that a constant power,
independent of $Y$, is inadequate.  Instead we write,
\beq
   f(A,Y,h) = Y^\beta A^{\alpha(Y)} f(\widetilde h).
\label{scaling}\eeq
and choose a particularly simple multifractal exponent with a linear
dependence on $Y$. The data corpus supports such a scaling law with
\beq
   \alpha = (0.76\pm0.02) + (0.10\pm0.01) Y,
\label{alpha}\eeq
with covariance $-0.984$, as shown in Figure \ref{fg.alpha}.  The data
from CERN-PS \cite{CERN-PS} was not used in this fit, because of the
large errors in this measurement; its inclusion does not change the
central values of the fit significantly but increases the errors. It would
be interesting to test in future experiments whether this exponent is
modified near $Y=0.1$ as the thresholds for $\psi$(2S) and the $\chi$s
are approached.

\begin{figure}[tp]
\begin{center}
\includegraphics[scale=0.7]{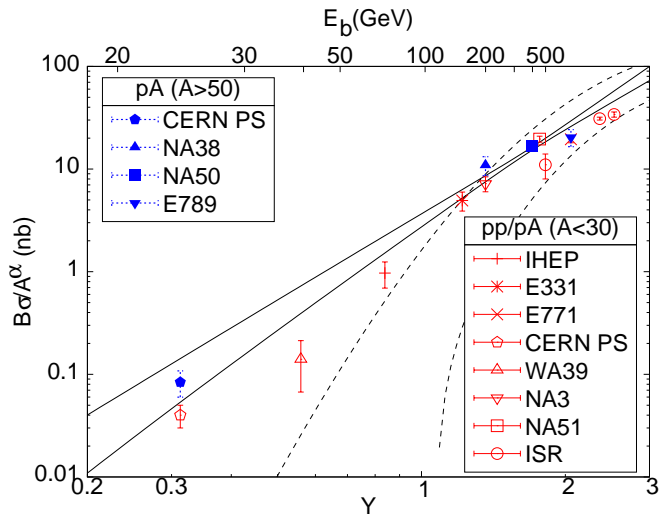}
\end{center}
\caption{Scaling plot for the experimental data on the total inclusive
 cross section for the production of $J/\psi$ in pA collisions. The band
 enclosed by full lines is the 68\% confidence limit of the two parameter fit
 of eq.\ (\ref{xfit}) to pA data for $A>50$, so excluding p-Be data from IHEP
 \cite{IHEP} and E331 \cite{E331} and p-Si data from E771 \cite{E771}.
 The band enclosed between the dashed curves is the 68\% confidence limit of
 a two parameter fit to the form $K[1-{\rm e}^{-Y}]^\nu$ to all pA data.}
\label{fg.xsec}\end{figure}

For a test of the scaling behaviour in eq.\ (\ref{scaling}) we constructed
the scaled cross section $B\sigma/A^\alpha$, using the largest $A$
from each experiment which had $A>50$. Fitting by a power of $Y$ gives
\beq
   Bf/S_0 = 3.2\pm0.5\;{\rm nb\/},\quad
   \beta = 3.0\pm0.3,
\label{xfit}\eeq
with covariance of $-0.970$.  In dimensionless units we have $f =
(3.4\pm0.5) \times 10^{-3}$.  The exponent is consistent with that
obtained in eq.\ (\ref{ppscale}). In fact, we have plotted the pp data
also in Figure \ref{fg.xsec} to show that the two sets of data are close,
but not identical. This is consistent with our expectation that the
multifractal exponent of $A$ is valid for large $A$.  It is clear from
Figure \ref{fg.xsec} that the data corpus does not provide a very
stringent test of scaling in near-threshold production cross sections.
Additional data would therefore be very useful.

It is interesting to extend the parametrization of eqs.\
(\ref{scaling},\ref{alpha},\ref{xfit}) to other resonances. The
exponent $\alpha$ for the ground state quarkonium is unlikely to
have a close relationship to that for higher resonances, since they
are more susceptible to decay due to medium effects. On the other
hand, one may ask whether there is any universality in exponents of
ground state quarkonia. To answer this, we examine the $\Upsilon$,
for which E772 \cite{E772} reported $\alpha=0.962\pm0.006\pm0.008$
at $\sqrt S=38.8$ GeV. Extrapolating the fit in eq.\ (\ref{alpha})
gives $\alpha=0.883\pm0.008$. To improve this we add a resonance
mass dependence: $\alpha = \alpha_0 + \alpha_1 Y + \alpha_h h$. The
fit of eq.\ (\ref{alpha}) along with the above measurement gives
$\alpha_h=0.012\pm0.002$, and $\alpha_0=0.64\pm0.02$. Interestingly
optical models of shadowing in very low-energy nucleon-nucleus scattering
predict $\alpha_0=2/3$. This value is in agreement with results for low
energy $\pi$ production \cite{pirho}. Coincidentally, the formula is
also in agreement with the measured value of $\alpha$ for $K$, $\rho$,
and $\omega$ production at low energy \cite{omega}.

It is interesting to compare the parametrization of eq.\ (\ref{scaling})
with other studies. The Glauber model treated in the eikonal approximation
has also been used in cold nuclear matter with the parametrization
\beq
 f(A,Y,h) = A\exp\left(-\gamma A^{1/3}\right) f(Y,h).
\label{glauber}\eeq
In the model, the dimensionless number $\gamma=\rho\sigma_{abs}\lambda$,
with $\rho$ being the nuclear density, $\sigma_{abs}$ having the
interpretation of a cross section for absorption of $J/\psi$ in
cold nuclear medium, and $\lambda A^{1/3}$ being the path length
of the $J/\psi$ in the nucleus (see, for example, \cite{carlos}).
Although the functional form $F(A)$ seems to be very different from
$A^\alpha$, they can be numerically close.  As a result, the Glauber model
cannot be experimentally distinguished from the simpler model of eq.\
(\ref{scaling}).  In fact, the data corpus can be used to parametrize
$\gamma = (0.27 \pm 0.02) - (0.092 \pm 0.007) Y$. Then, using the values
$\rho=0.16$/fm$^3$ and $\lambda=1.1$ fm, we can write
\beq
 \sigma_{abs}=(15\pm1)-(5.2\pm0.4) Y {\rm\ mb}, 
    \qquad{\rm for\ } Y\le2.5.
\label{sabs}\eeq
This is consistent with the values extracted in \cite{NA60}.

Different forms have also been used for the scaling function
$f(Y,h)=S_0\sigma A^{-\alpha}$. A parametrization suggested in
\cite{linnyk} is compatible with the data for $Y>1$, but differs
from that of eq.\ (\ref{scaling}) near threshold.  A parametrization
of the threshold effect as $(1-1/\exp Y)^\nu$, as in \cite{E771}, is
compatible with data for $Y>1$, but differs from eq.\ (\ref{scaling})
near the threshold, as shown in Figure \ref{fg.xsec}.  Fermi motion
based explanations of the discrepancy with data for $Y<1$ are ruled out
because two pieces of the data come from pp experiments.

We argue that the crossover between these forms is physical and
interesting.  Near threshold $Y\simeq(1-1/\exp Y)$, so the powers
$\beta$ and $\nu$ can be compared. The fit value $\beta\simeq3$ can be
interpreted within the old ``spectator counting rules'' \cite{farrar},
as indicating that there are 2 spectators. Since hidden charm production
is an OZI-violating process, this is consistent with a purely hadronic
origin of near threshold processes. On the other hand, the best fit
value of $\nu\simeq11$ (Figure \ref{fg.xsec}) implies that there are
6 spectators. This is consistent with counting the valence quarks from
two participating hadrons in a process dominated by gluon fusion. So the
crossover from one regime to another may give us a clue to the regime
of validity of perturbative QCD.

The scaling theory is unable to relate the values of $f(\widetilde h)$ in
experiments with different initial states: for example pA, $\overline{\rm
p}$A, $\pi^\pm$A, \etc. If some form of the factorization theorems were valid,
then there could be relations such as
\beq
   \frac{f(pA\to H)}{f(pA'\to H)} = \frac{f(\pi A\to H)}{f(\pi A'\to H)},
\label{factor}\eeq
where $f(ab\to h)=\sigma S_0$ and $\sigma$ is the total inclusive
scattering cross section for the reaction $ab\to h$ at a fixed $Y$. Such
tests could also be performed with light ions instead of pions. The
validity of factorization would be an interesting investigation especially
in view of the fact that the connection between hadron decays and the
CKM matrix elements can only be made with this assumption near threshold.

Proceeding beyond total inclusive cross sections near threshold is hard
due to the paucity of data.  There is scattered evidence for cold nuclear
effects in $\langle p_\T^2\rangle$, from NA38 \cite{NA38} and HERA-B
\cite{HERAB}.  However, there is too little data for a systematic study
of the effect. The sparse corpus shows a roughly linear rise of $\langle
p_\T^2\rangle$ with $Y$, from a vanishing value at $Y=0$.  Clearly, high
statistics studies of $p_\T$ and $x_\F$ distributions of the $J/\psi$
near threshold would be very welcome.

Since data with beam energy $E_b<100$ GeV is very sparse
\cite{CERN-PS,WA39,Aubert},  the SIS-100 accelerator at GSI presents an
opportunity to probe the region of $Y\le0.4$ very thoroughly with modern
statistics. With a beam luminosity of 1 Hz/nb, fair event rates could be
obtained, and these scaling laws can be tested well. This would make the
GSI an ideal test bed for exploring the near-threshold production process
for $J/\psi$ as well as cold nuclear effects, including questions about
factorization and $p_\T$ and $x_\F$ distributions. A wish list would
contain measurements with pp and a variety of pA collisions to check the
scaling of eq.\ (\ref{alpha}). The pp data could also be used to check
the scaling exponent of eq.\ (\ref{ppscale}) and whether it is compatible
with the pA result of eq.\ (\ref{xfit}). A range of A can be used to
test the region of validity of the power law $A^\alpha$. A systematic
study of $p_\T$ and $x_\F$ distributions would also be extremely useful.

In summary, we have extracted a power law parametrization of the cross
section for near-threshold $J/\psi$ production off cold nuclear matter.
The results are given in eqs.\ (\ref{scaling},\ref{alpha},\ref{xfit}).
Such power laws are more than just a parametrization, since they reveal
certain dynamical symmetries of hadronic systems, which equate a physical
system with one $Y$ and $A$ to another 
with different $Y$ and $A$. These identities constitute renormalization
group transformations, and should eventually be computable from
QCD. Interestingly, it seems that the multifractal exponent $\alpha$ of
eq.\ (\ref{alpha}) can be extended to the production of the $\Upsilon$ via
\beqa
\nonumber
  \alpha(Y,h) &=& (0.64\pm0.02) + (0.10\pm0.01)Y \\
     &&\qquad + (0.012\pm0.002) h.
\label{universal}\eeqa
Coincidentally, this form it also reproduces the exponent required for
inclusive $\pi$, $K$, $\rho$ and $\omega$ production, but not for $\psi'$
or $\phi$ production.  These scaling laws present fundamental restrictions
on QCD, and therefore should be of priority in upcoming low-energy and
high-intensity experiments at the SIS-100/300.

PPB would like to thank TIFR for local hospitality during a visit in
which this work was done, and Dr.\ Subhasis Chattopadhyay for travel
funding from his DAE-SRC award under the scheme No. 2008/21/07-BRNS/2738.

 \end{document}